\documentclass[conference,a4paper]{IEEEtran}
\usepackage{float}
\usepackage{amsmath}
\usepackage{amsthm}
\usepackage{amssymb}	
\usepackage{graphicx}
\usepackage{subfigure}
\usepackage{enumitem}
\usepackage{algorithm}
\usepackage{stfloats}
\usepackage{cite}
\usepackage{cuted}
\usepackage{color}
\usepackage[margin=0.7in]{geometry}
\newcommand*{\J}{\jmath}%
\DeclareMathOperator{\sinc}{sinc}
\usepackage{amsmath,amssymb,mathtools,bm,etoolbox}

\newtheorem{my_lemma}{Lemma}
\newtheorem{my_corollary}{Corollary}
\newtheorem{my_proposition}{Proposition}

\usepackage{color}
\usepackage{mathtools}
\usepackage{suffix}
\usepackage{breqn}
\usepackage{hhline}
\usepackage{cite}
\DeclarePairedDelimiterXPP\Aver[1]{\mathbb{E}}{[}{]}{}{
	
	#1
}
\DeclarePairedDelimiterX\MeijerM[3]{\lparen}{\rparen}%
{\,#3\delimsize\vert\begin{matrix}#1 \\ #2\end{matrix}}
\newcommand\MeijerG[8][]{%
	G^{\,#2,#3}_{#4,#5}\MeijerM[#1]{#6}{#7}{#8}}

\WithSuffix\newcommand\MeijerG*[7]{%
	G^{\,#1,#2}_{#3,#4}\MeijerM*{#5}{#6}{#7}}

\title{Path Loss Modeling for RIS-Assisted Wireless System with Direct Link and Elevation Factors} 
\author{
	\IEEEauthorblockN{ Vinay Kumar Chapala, Pratham Sharma, Sameer Sharma, and S.~M.~Zafaruddin}\\
	\IEEEauthorblockA{ Department of Electrical and Electronics Engineering, 
		BITS Pilani, Pilani Campus, Pilani-333031, Rajasthan, India
		\\ Email: \{syed.zafaruddin\}@pilani.bits-pilani.ac.in}
	
	\thanks{This work was supported in part by the Science and
		Engineering Research Board (SERB), Department of Science and Technology
		(DST), Government of India, under MATRICS Grant MTR/2021/000890 and Core Research Grant CRG/2023/008040.}
}

\IEEEoverridecommandlockouts
\begin{document}
	
	\maketitle
	\begin{abstract}
The present path loss models for wireless systems employing reconfigurable intelligent surfaces (RIS) do not account for the elevation of the transmitter, receiver, and RIS module.
In this paper, we develop an analytical model for path loss of  a wireless system utilizing an $N\times M$-element RIS module positioned above the ground surface with elevated transmitter and receiver configurations. Furthermore, we integrate the direct link into the path loss model to enhance its applicability, a crucial aspect often neglected in previous research. We also present simplified analytical expressions for path loss under various configurations, including near-field and far-field scenarios. These expressions elucidate the impact of elevation factors on path loss, facilitating more accurate signal quality estimation at the receiver. Simulation results corroborate that accounting for elevated RIS modules and transceiver units can yield improved deployment strategies for RIS-based wireless systems.
	\end{abstract}

	\section{Introduction}
 Reconfigurable intelligent surface (RIS) based wireless system addresses the challenge of controlling the unpredictable nature of the radio channel by dynamically reflecting electromagnetic waves to enhance transmission quality \cite{Basar_2019,Chongwen_2019,Wu_Qing_2020,Ming_Zheng_2020,Yishi_2021}. The RIS is composed of multiple passive reflective elements, which serve to counteract the multiplicative effects of the channel by coherently redirecting signals to a designated destination \cite{Ming_Zheng_2020,Yishi_2021}. An accurate understanding of the signal strength at the receiver is pivotal for optimizing the performance, efficiency, and reliability of wireless communications. Path loss modeling is increasingly crucial for RIS-assisted wireless systems to optimize deployment scenarios due to the increased complexity of signal attenuation compared to direct links.

There has been a growing research focus on the development of analytical and parametric path loss models for RIS-assisted wireless systems 	
\cite{Ozdogan_2020, Danufane_2020, Garcia_2020, Najafi_2020, Tang_2021,Wankai_Tang_2021, Danufane_2021, Steven_W_2021, Yu_2021, Li_Sifeng_2021, Huang2022Proc,Degli_Esposti_2022, Wang_Jihong_2023}. The authors in \cite{Ozdogan_2020}  employed physical optics techniques to derive an analytical expression for the scattered field, enabling the modeling of path loss for a RIS configured to redirect incoming waves from a far-field source towards a receiver also positioned in the far-field. They demonstrated that the received signal power scales proportionally with the square of the RIS area and inversely with the product of the distance from the transmitter to the RIS and from the RIS to the receiver. In their study, Danufane \emph{et al.}  \cite{Danufane_2020}  utilized the general scalar theory of diffraction to derive approximate closed-form expressions for the electric field reflected by RIS in both short and long transmission distance regimes, elucidating the conditions under which these surfaces transform into mirrors, RIS, or passive reflecting beamformers.

In the seminal work,  Tang \emph{et al.} \cite{Tang_2021} developed free-space path loss models for RIS-assisted wireless communications for different scenarios by studying the physics and electromagnetic nature of RISs. The proposed models, which were first validated through extensive simulation results, revealed the relationships between the free-space path loss of RIS-assisted wireless communications and the distances from the transmitter/receiver to the RIS, the size of the RIS, the near-field/far-field effects of the RIS, and the radiation patterns of antennas and unit cells. In addition, three fabricated RISs were utilized to further corroborate the theoretical findings through experimental measurements conducted in a microwave anechoic chamber. The paper  \cite{Wankai_Tang_2021} further refined the model presented in the previous work \cite{Tang_2021}, enhancing its depiction of the influence of transmitter, receiver, and RIS unit cells' directivity on path loss. The path-loss model proposed in the study \cite{Danufane_2021} utilized the vector generalization of Green's theorem and computable integral dependent on transmission distances, radio wave polarization, RIS size, and desired surface transformations. Closed-form expressions were derived for two asymptotic regimes representing far-field and near-field transmission, allowing for unveiling the impact of various design parameters and a discussion of the differences and similarities between the two regimes. The author in \cite{Steven_W_2021} devised a technique to compute path loss for RIS by considering an array of parameterizable element patterns and extending it to a continuous electromagnetic surface. The authors in \cite{Li_Sifeng_2021} conducted simulations to observe the impact of multipath propagation on the RIS-assisted communication system model, supplementing the free-space path loss model.

A recent comprehensive review of channel characterization and modeling for RIS-assisted wireless systems was presented in \cite{Huang2022Proc}. Furthermore, in \cite{Wang_Jihong_2023}, the authors introduced path loss models for a double RIS system, building upon the groundwork laid in \cite{Tang_2021}. However, existing path loss models for RIS-assisted systems have not considered the elevation of the transmitter, receiver, and RIS module. Notably, path loss models for direct links already incorporate the elevation of the transmitter and receiver, as outlined in extensive research and embraced by standards such as \cite{ETSI5G}. Typically, RIS is installed at an elevated location, underscoring the necessity of incorporating RIS elevation into the path loss model. 
Inaccuracies in path loss modeling may result in suboptimal system performance, diminished communication quality, reduced coverage, and heightened interference levels.

This paper introduces an analytical model for path loss in a wireless system incorporating an $N\times M$-element RIS module positioned above the ground, with elevated transmitter and receiver setups. Further, we incorporate the direct link into the path loss model, a critical consideration often overlooked in prior studies. We also provide simplified analytical formulas for path loss under various setups, including near-field and far-field scenarios. These expressions highlight the influence of elevation factors on path loss, enabling a more precise estimation of signal quality at the receiver. Simulation results affirm that accounting for elevated RIS modules and transceiver units can enhance deployment strategies for RIS-based wireless systems through accurate path loss estimation.

  \begin{figure}
	\centering
	{\includegraphics[scale=0.4]{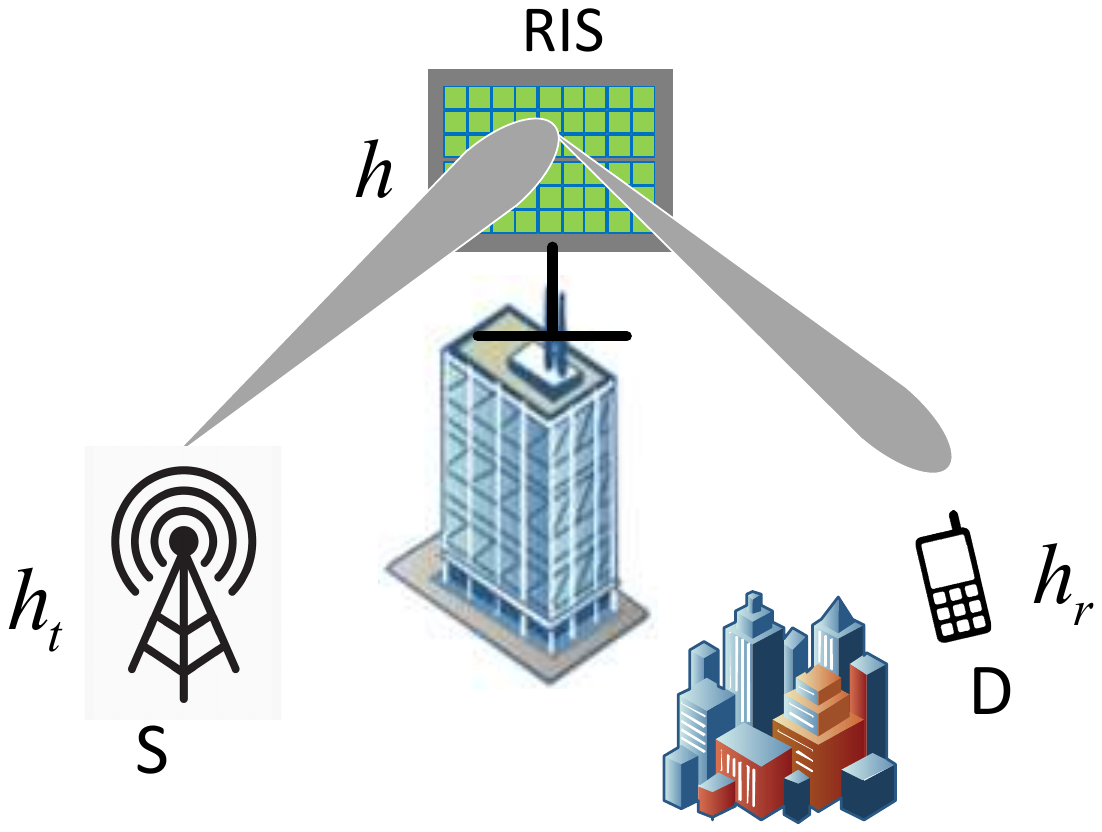}} 
	\vspace{-3cm}
	\caption{RIS-assisted system with an elevated transceiver and RIS mounted at a height.}
	\label{fig:two_ray_ris}
\end{figure}

\begin{table*}[tp] 
	\caption{{List of Parameters with Description}}
	\label{table1}
	\vspace{-.4cm}
	\begin{center}
		\begin{tabular}{ | c | c| c | c|} 
			\hline
			\textbf{Parameters} & \textbf{Description} & \textbf{Parameters} & \textbf{Description}\\ 
			\hline
			$N$ & Number of unit cells in a row & $M$ & Number of unit cells in a column \\ 
			\hline
			$h_r$ & Elevation of the receiver & $h_t$ & Elevation of the transmitter \\ 
			\hline
			$d$ & Distance between transmitter and receiver & $d_1$ & Distance between RIS and transmitter \\ 
			\hline
			$d_2$ & Distance between RIS and receiver & $d_l$ & Direct link distance \\ 
			\hline
			$h$ & Elevation of the RIS & $G$ & Attenuation at the RIS element \\
			\hline
			$F_{direct}^{combine}$ & Combined field radiation pattern in the LOS & $F_{n,m}^{combine}$ & Combined field radiation pattern in the reflected path \\ 
			\hline
			$G_t$ & Gain of the transmit antenna & $G_r$ & Gain of receive antenna\\ 
			\hline
			$r_{n,m}^{t}$ & Distances from the transmitter to the RIS element & $r_{n,m}^{r}$  
			& Distances from the RIS element to the receiver   \\ 
			\hline
			$\theta_{n,m}^{t}$ & Elevation angle from the RIS to the transmit antenna & $\psi_{n,m}^{t}$ & Azimuth angles from the RIS to the transmit antenna\\ 
			\hline
			$\theta_{n,m}^{r}$ & Elevation angle from the RIS to the receive antenna & $\psi_{n,m}^{r}$ & Azimuth angles from the RIS to the receive antenna  \\ 
			\hline
			$U_{n,m}$& Represents Unit cell & $d_{x},d_{y}$	& Dimensions of unit cell \\
			\hline
			$A_{n,m}$ & Controllable amplitude & $\phi_{n,m}$ & Phase of RIS element \\ 
			\hline
			$\lambda$ & Wavelength of propagation & $\Delta\phi$ & Phase difference between LOS and reflected signal\\ 
			\hline
			$sinc(x)$ &  $\frac{sin({\pi}x)}{{\pi}x}$ & $\Gamma$ & Reflection coefficient \\ 
			\hline
		\end{tabular}
		
	\end{center}
\end{table*}

\section{System Model}

Consider a wireless communication scenario involving a source and destination aided by a RIS, as illustrated in Fig.~\ref{fig:two_ray_ris}. In this setup, there is a transmitter at height $h_{t}$, a RIS with dimensions $N\times M$ located at height $h$ (where $N$ and $M$ are the numbers of unit cells in the RIS's rows and columns, respectively), and a receiver positioned at height $h_{r}$. A direct connection is assumed, enabling a generalized path loss model. When the RIS is positioned at a specific height $h$, its origin and geometric center align. The distance between the transmitter and receiver is represented as $d$, while $d_1$ and $d_2$ denote the distances between the transmitter and the RIS element and the RIS and the receiver, respectively. Further, $\theta$ signifies the angle between the transmitter and the RIS element.

Let $F_{\text{direct}}^{\text{combine}}$ represent the combined field radiation pattern in the line-of-sight (LOS) direction derived from the transmit and receive antenna patterns. Moreover, $F_{n,m}^{\text{combine}} = F_{\text{tx}}(\theta_{n,m}^{\text{tx}},\psi_{n,m}^{\text{tx}}) F_{\text{}}^{}(\theta_{n,m}^{t},\psi_{n,m}^{t}) F_{\text{}}^{}(\theta_{n,m}^{r},\psi_{n,m}^{r}) F_{\text{rx}}(\theta_{n,m}^{\text{rx}},\psi_{n,m}^{\text{rx}})$ incorporates the normalized radiation pattern's impact. In this context, $G_{t}$ and $G_{r}$ denote the gains of the transmit and receive antennas, respectively, while $G$ represents the attenuation at the RIS element $U_{n,m}$. Variables $r_{n,m}^{\text{t}}$, $r_{n,m}^{\text{r}}$, $\theta_{n,m}^{\text{t}}$, $\psi_{n,m}^{\text{t}}$, $\theta_{n,m}^{\text{r}}$, and $\psi_{n,m}^{\text{r}}$ indicate the distances from the transmitter to the RIS element, from the RIS element to the receiver, and the elevation and azimuth angles from the RIS element to the transmitter and receiver, respectively.

Further, $\theta_{n,m}^{\text{tx}}$, $\psi_{n,m}^{\text{tx}}$, $\theta_{n,m}^{\text{rx}}$, and $\psi_{n,m}^{\text{rx}}$ denote the elevation and azimuth angles from the transmit antenna to the RIS element and from the receive antenna to the RIS element. Here, $U_{n,m}$ represents the unit cell, and $dx$ and $dy$ denote the dimensions of the unit cell, typically within the range of $\frac{\lambda}{10}$ to $\frac{\lambda}{2}$. Furthermore, $\Gamma_{n,m} = A_{n,m} e^{\phi_{n,m}}$ represents the reflection coefficient of the RIS cell, where $A_{n,m}$ and $\phi_{n,m}$ represent the controllable amplitude and phase of the RIS element, respectively. Table~\ref{table1} describes the parameters for a quick reference.

\section{Path Loss Modeling for RIS-Assisted Wireless with Elevation Factors}
This section aims to develop analytical models that delineate the path loss in a wireless system enhanced by RIS. Through systematically examining propagation traits and interaction dynamics among transmitted signals, RIS components, and the receiver, we aim to furnish holistic frameworks for assessing the signal attenuation encountered during its propagation across the system.

It is essential to define and express the distances and heights involved in the system, as formulated in the following Proposition:
\begin{my_proposition}
	If $h_t$ represents the height of the transmit antenna, $h_r$ signifies the height of the receiver antenna, and $h$ indicates the height of the RIS above the ground surface, then the direct link distance between the transmitter and receiver can be expressed as: 
	\begin{eqnarray}\label{eq:dl}
	d_{l} = \sqrt{(h_{t}-h_{r})^{2}+d^{2}}
	\end{eqnarray}
	Further, the phase difference between the line-of-sight (LOS) and reflected signal asymptotically follows:
	\begin{equation}\label{eq:phase}
	\Delta\phi = \frac{4\pi \big[h^{2}+h_{t}h_{r}-h h_{t}-h h_{r}\big]}{\lambda d}
	\end{equation} 
	where $d$ represents the distance between the transmitter and receiver, and $\lambda$ denotes the wavelength of propagation.
\end{my_proposition}
\begin{IEEEproof}
	We revisit the system model in Fig.~\ref{fig:two_ray_ris} to enhance our understanding of the interplay between different distances and heights, as illustrated in Fig.~\ref{fig:Two_Ray_2}. This updated depiction will allow us to analyze the relationships among these parameters more accurately.
	\begin{figure}
		\centering
		{\includegraphics[scale=0.4]{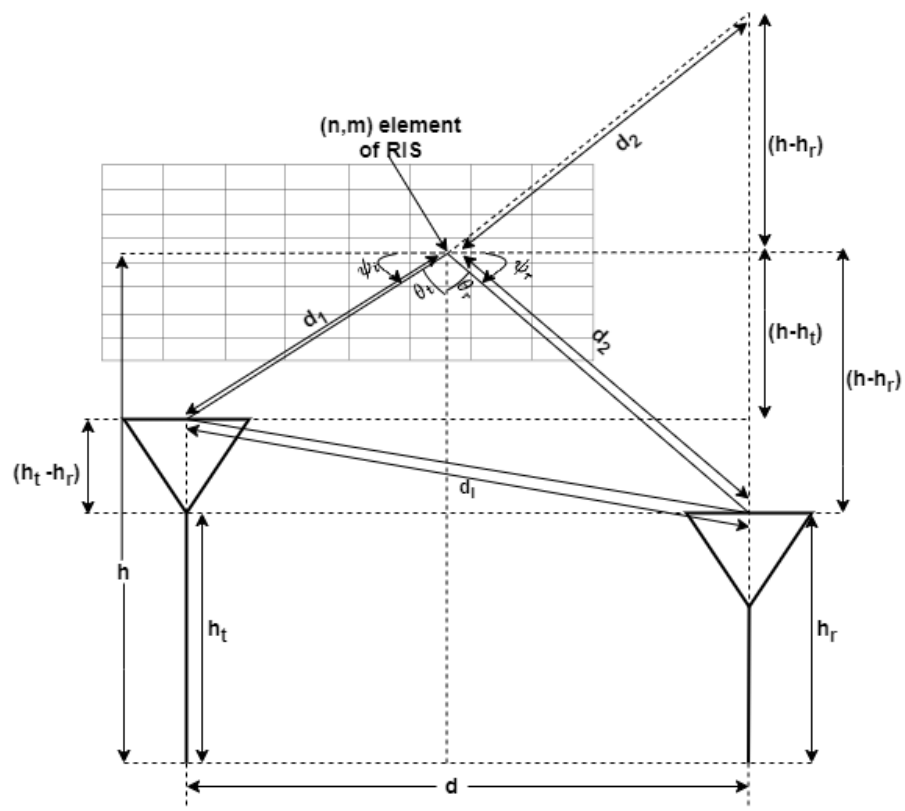}}
		\caption{Illustration depicting the geometric configuration of transceiver heights and the elevation of RIS-assisted  wireless system.}
		\label{fig:Two_Ray_2}
	\end{figure}	
	Utilizing geometric visualization, it becomes apparent that $d_{l} = \sqrt{(h_{t}-h_{r})^{2}+d^{2}}$, validating \eqref{eq:dl}. To establish \eqref{eq:phase}, we start by expressing $d_{1}+d_{2} = \sqrt{(h-h_{t}+h-h_{r})^{2}+d^{2}}$, which simplifies to $d_{1}+d_{2} =\sqrt{(2h-h_{t}-h_{r})^{2}+d^{2}}$. Assuming $d\gg 2h-h_t-h_r$ and applying Taylor's series, we approximate the path difference as $\Delta l= d_{1}+d_{2}-d_{l} \approx \frac{2}{d}\big[h^{2}+h_{t}h_{r}-h h_{t}-h h_{r}\big] $. Finally, obtaining the phase difference $\Delta \phi=\frac{2 \pi \Delta l}{\lambda}$ is straightforward, as presented in \eqref{eq:phase}.
\end{IEEEproof}

In what follows, we construct analytical models to examine received power within the comprehensive framework encompassing direct links and signals received via RIS elements.
\begin{my_lemma}
	The received signal power in RIS-assisted wireless communications with elevation factors and a direct link is given as
	\begin{flalign}\label{eq:main}
	P_r = P_t \bigg[\frac{\lambda}{4\pi}\bigg]^2\bigg|\frac{\sqrt{G_{t}G_{r}}\sqrt{F_{direct}^{combine}}}{d_{l}} + \frac{\sqrt{G_r G_t G d_x d_y}}{2 \sqrt{\pi}} \nonumber \\ \sum_{m=1 - M/2}^{M/2} \sum_{n=1 - N/2}^{N/2} \frac{\sqrt{F_{n,m}^{combine}} \Gamma_{n,m}}{r_{n,m}^{t} r_{n,m}^{r}} e^{\frac{-j2\pi(r_{n,m}^{t} + r_{n,m}^{r} - d_{l})}{\lambda}}\bigg| ^2 
	\end{flalign}
	where $d_{l} = \sqrt{(h_{t}-h_{r})^{2}+d^{2}}$
\end{my_lemma}
\begin{IEEEproof}
	Assuming a single-element RIS, we derive the expression for received power utilizing the widely employed two-ray model \cite{andreabook}:
	\begin{equation}\label{eq:two_ray_1_1}
	P_{r} = P_{t} \bigg[\frac{\lambda}{4\pi}\bigg]^{2} \bigg| \frac{\sqrt{G_{a}G_{b}}}{d_l} + \frac{\Gamma \sqrt{G_{c}G_{d}}e^{-\J\Delta\phi}}{d_1+d_2} \bigg|^{2}
	\end{equation}
	Here, $G_{a}$ and $G_{b}$ represent the transmit and receiver field radiation patterns in the LOS direction, while $G_{c}$ and $G_{d}$ pertain to the reflected path. $\Gamma$ denotes the reflection coefficient, and $\Delta\phi=\frac{2\pi(d_1+d_2-d_l)}{\lambda}$ signifies the phase difference between the two received signal components. Using Fig.~\ref{fig:Two_Ray_2}, the coordinates of RIS element $U_{n,m}$ are given as  $((m-\frac{1}{2})d_{x},(n-\frac{1}{2})d_{y},h)$,  and the distance from the transmit antenna at $(x,y,z)$ to the RIS element $U_{n,m}$ is $r_{n,m}^{t}=\sqrt{(x-(m-\frac{1}{2})d_{x})^{2}+(y-(n-\frac{1}{2})d_{y})^{2}+(z-h)^{2}}$. Similarly, distance $r_{n,m}^{r}$ from RIS to the receiver can be obtained. Leveraging the approach from \cite{Tang_2021}, we extend our analysis to incorporate the contributions from all reflecting elements by integrating these reflections along with the direct link to get \eqref{eq:main}.
\end{IEEEproof}
The expression in \eqref{eq:main} represents a generalization of the path loss model for an RIS with $MN$ elements, considering the elevation of both the RIS and transceiver units, including the direct link.

In the scenario of far-field beamforming, where the distances between the transmitter, receiver, and reflecting elements are significantly large compared to the wavelength, the angular parameters $(\theta_{n,m}^{t},\psi_{n,m}^{t})$ and $(\theta_{n,m}^{r},\psi_{n,m}^{r})$ can be effectively approximated as $(\theta_{t},\psi_{t})$ and $(\theta_{r},\psi_{r})$, respectively. Here, $\theta_{t}$ and $\theta_{r}$ represent the azimuth angles of the transmitter and receiver, while $\psi_{t}$ and $\psi_{r}$ denote the elevation angles.

Furthermore, in the far-field regime, the radiation patterns at the transmitter and receiver antennas, denoted by $F_{}^{tx}(\theta_{n,m}^{tx},\psi_{n,m}^{tx})$ and $F_{}^{rx}(\theta_{n,m}^{rx},\psi_{n,m}^{rx})$, respectively, can be approximated as unity, effectively simplifying to $F_{}^{tx}(\theta_{n,m}^{tx},\psi_{n,m}^{tx})\approx 1$ and $F_{}^{rx}(\theta_{n,m}^{rx},\psi_{n,m}^{rx})\approx1$. Similarly, for the direct link, $F_{direct}^{combine}\approx1$.

\begin{my_corollary}
	The received signal power in RIS-assisted wireless communications with elevation factors and direct link under far-field is given as
	\begin{flalign}\label{eq:far_field}
	P_r &= P_t \big[\frac{\lambda}{4\pi}\big]^2\Bigg|\frac{\sqrt{G_{t}G_{r}}}{d_{l}} \big[1+j\frac{4\pi}{\lambda d} \big[h^{2}+h_{t}h_{r}-h h_{t}-h h_{r}\big]\big] \nonumber \\&+ M N Ae^{\J\phi}\frac{\sqrt{G_r G_t G F_{}^{}(\theta_{t},\psi_{t})F_{}^{}(\theta_{r},\psi_{r}) d_x d_y}}{2 \sqrt{\pi}d_{1} d_{2}}\nonumber\\&e^{-\J\frac{2\pi}{\lambda}((h-h_{t})\cos{\theta_{t}}+(h-h_{r})\cos{\theta_{r}})}  \nonumber\\&\frac{\sinc(\frac{M\pi}{\lambda}(\sin{\theta_{t}}\cos{\psi_{t}}+\sin{\theta_{r}}\cos{\psi_{r}})d_{x})}{\sinc(\frac{\pi}{\lambda}(\sin{\theta_{t}}\cos{\psi_{t}}+\sin{\theta_{r}}\cos{\psi_{r}})d_{x})}\nonumber\\&\frac{\sinc(\frac{N\pi}{\lambda}(\sin{\theta_{t}}\sin{\psi_{t}}+\sin{\theta_{r}}\sin{\psi_{r}})d_{y})}{\sinc(\frac{\pi}{\lambda}(\sin{\theta_{t}}\sin{\psi_{t}}+\sin{\theta_{r}}\sin{\psi_{r}})d_{y})} \Bigg| ^2 
	\end{flalign}
\end{my_corollary}
\begin{IEEEproof}
	The proof is presented in Appendix A.
\end{IEEEproof}
It can be seen that  \eqref{eq:far_field} provides a detailed representation of the received power for a RIS-assisted system in the far-field scenario, a typical scenario for wireless communication.

We can examine a specific scenario that maximizes the received power in the far-field case. This situation occurs when $\sin{\theta_{t}}\cos{\psi_{t}}+\sin{\theta_{r}}\cos{\psi_{r}}=0$ and $\sin{\theta_{t}}\sin{\psi_{t}}+\sin{\theta_{r}}\sin{\psi_{r}}=0$, indicating $\theta_{t}=\theta_{r}$ and $\psi_{t}=\psi_{r}+\pi$. Substituting these in  \eqref{eq:far_field}, the maximum received power for the far field case can  is expressed as:
\begin{flalign}\label{eq:far_max}
P_r^{max}& = P_t \bigg[\frac{\lambda}{4\pi}\bigg]^2\bigg|\frac{\sqrt{G_{t}G_{r}}}{d_{l}} +\nonumber \\& M N Ae^{\J\phi}\frac{\sqrt{G_r G_t G F_{}^{}(\theta_{t},\psi_{t})F_{}^{}(\theta_{r},\psi_{r}) d_x d_y}}{2 \sqrt{\pi}d_{1} d_{2}} \bigg|^2 \nonumber \\&+ P_t \frac{G_{t}G_{r}}{d_{l}^{2}d^{2}} \bigg[h^{2}+h_{t}h_{r}-h h_{t}-h h_{r}\bigg]^2
\end{flalign}
The received power in \eqref{eq:far_max} provides a simple formulation encompassing various parameters for efficient computation.
Assuming a single-element RIS, we can further simplify the received power in the far-field case as
\begin{equation}\label{eq:general}
P_{r} = P_{t} \big[\frac{\lambda}{4\pi}\big]^{2} \left| \frac{\sqrt{G_{t}G_{r}}}{d_{l}} + \frac{\Gamma\sqrt{G_{t}G_{r}}e^{-\J\Delta\phi}}{d_{1}+d_{2}} \right|^{2}
\end{equation}
Applying the results of Proposition 1 in \eqref{eq:general}, we get:
\begin{equation}\label{eq:single}
P_{r} = P_{t} \bigg[\frac{\sqrt{G_{t}G_{r}}\big[h^{2}+h_{t}h_{r}-h h_{t}-h h_{r}\big]}{d^{2}}\bigg]^{2}
\end{equation}
Finally, we consider the near-field scenario, where either the transmitter or receiver is very close to the RIS unit. In the following Corollary, we derive the received power:
\begin{my_corollary}
	The received signal power in RIS-assisted wireless communications with elevation factors and direct link under near-field broadcasting is given as
	\begin{flalign}\label{eq:two_ray_ris_16}
	P_r &= P_t \big[\frac{\lambda}{4\pi}\big]^2\bigg|\frac{\sqrt{G_{t}G_{r}}\sqrt{F_{direct}^{combine}}}{d_{l}} +Ae^{\J\phi}\frac{\sqrt{G_r G_t}}{d_{1}+d_{2}}\bigg| ^2\nonumber \\&+ P_tA^{2}\frac{G_r G_t}{(d_{1}+d_{2})^{2}d^{2}}\big[h^{2}+h_{t}h_{r}-h h_{t}-h h_{r}\big]^2 
	\end{flalign}
\end{my_corollary}
\begin{IEEEproof}
	The proof is presented in Appendix B.
\end{IEEEproof}
\begin{figure*}[tp]
	\centering
	\subfigure[Effect of phase compensation $\phi$ on RIS with direct link.]{\includegraphics[scale=0.35]{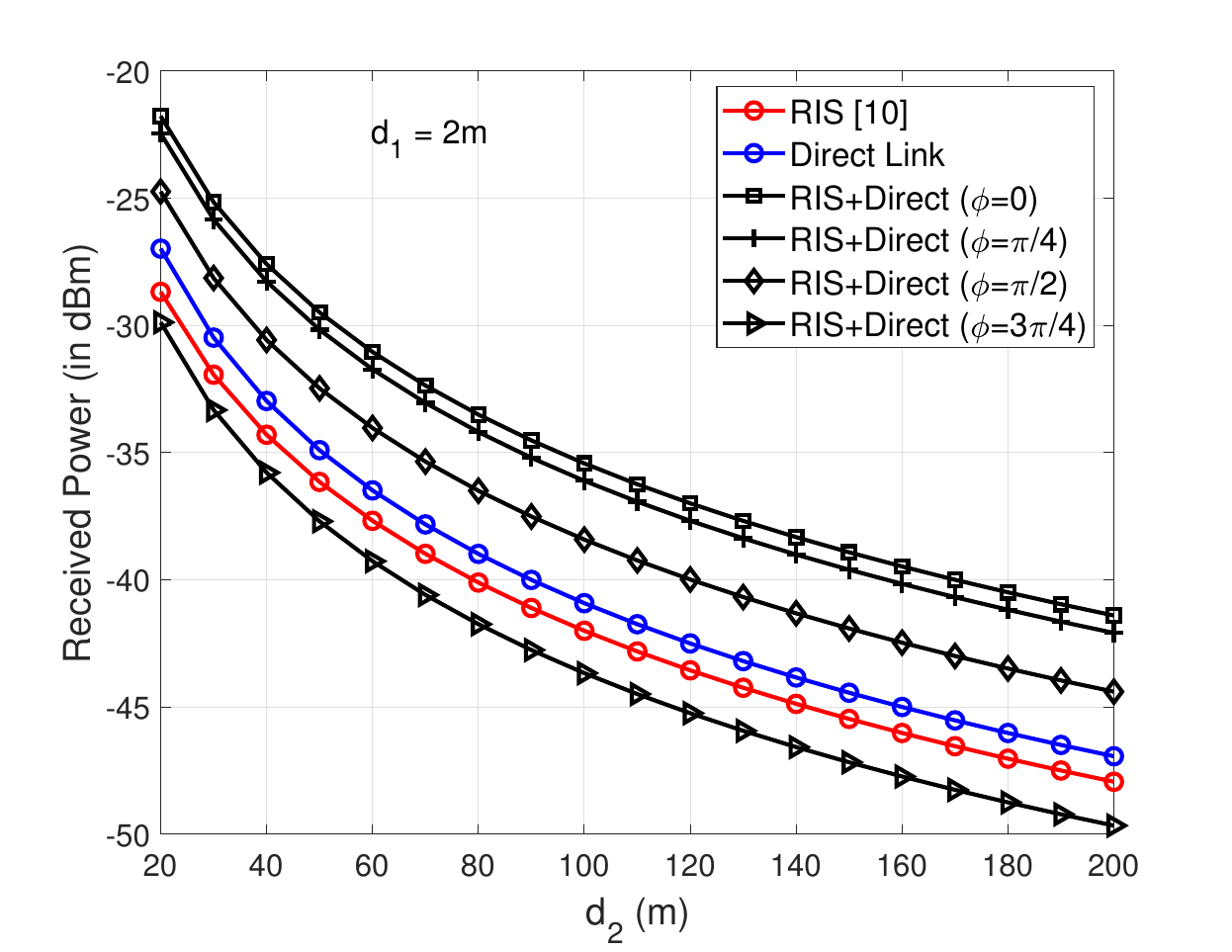}} \hspace{-2mm}
	\subfigure[Effect of RIS elevation factor $h$.]{\includegraphics[scale=0.35]{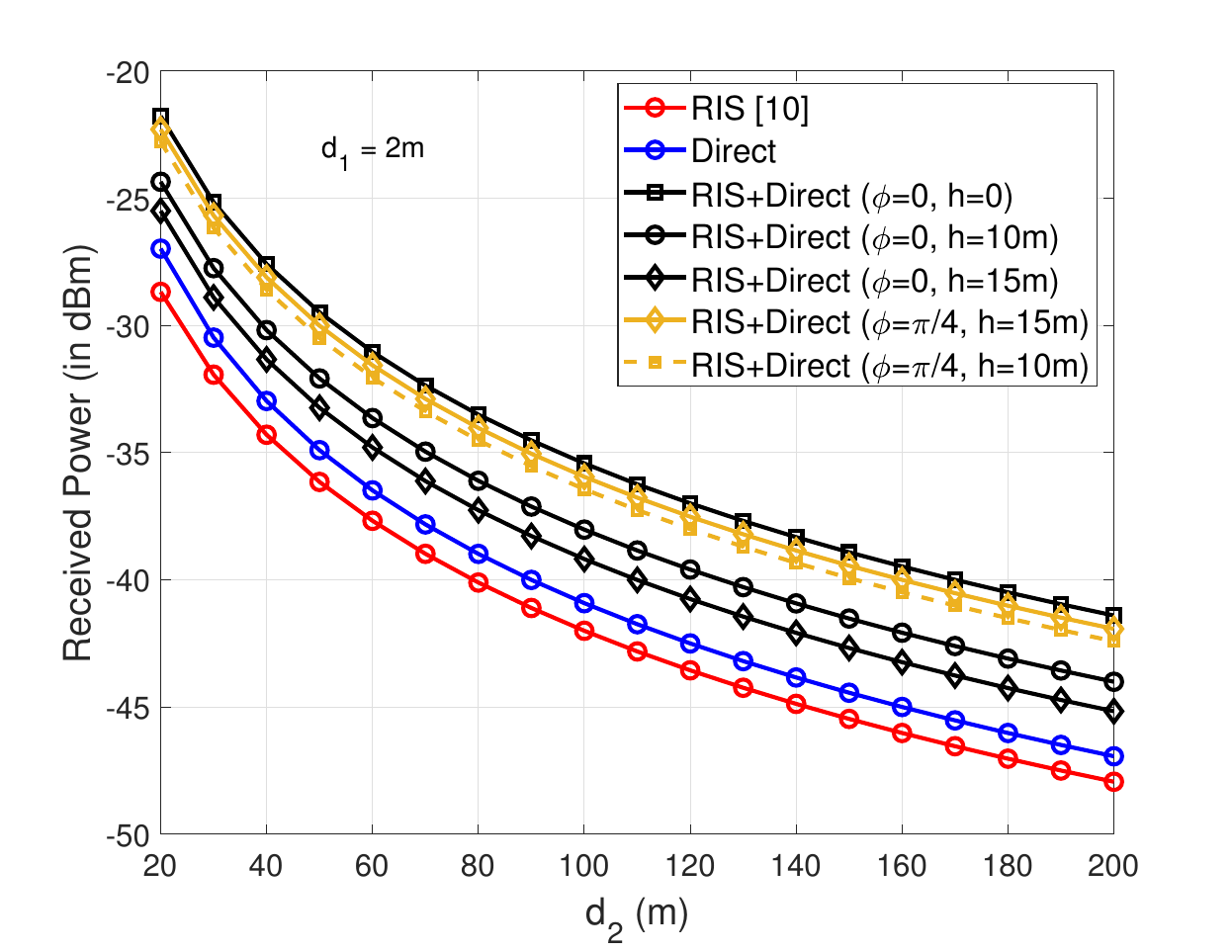}} \hspace{-2mm}
	\caption{Received power (dBm) for near-field broadcasting scenario with a transmit power of $10$\mbox{dBm}. }
	\label{fig:NF_BF}
\end{figure*}

\section{Numerical Results}\label{sec:sim_num_res}
In this section, we employ MATLAB-based simulations\footnote{ All simulation codes are available in the Github repository: https://github.com/prathamsharmaBITS/RIS-pathloss-modelling} to confirm the validity of our analytical models and illustrate how the elevation factors of the transceiver and RIS affect the received signal power, both with and without a direct link. Experimental validation will be part of the expanded version of the paper. Nonetheless, we have maintained identical simulation parameters to those used in \cite{Tang_2021} to ensure a fair comparison and a better appreciation of the proposed methodology. Numerical simulations are conducted using the general formula in \eqref{eq:main}, far-field in  \eqref{eq:far_field}, and the near-field broadcasting in \eqref{eq:two_ray_ris_16}.

\begin{figure*}[tp]
	\centering
	\subfigure[Effect of RIS elevation factor $h$.]{\includegraphics[scale=0.35]{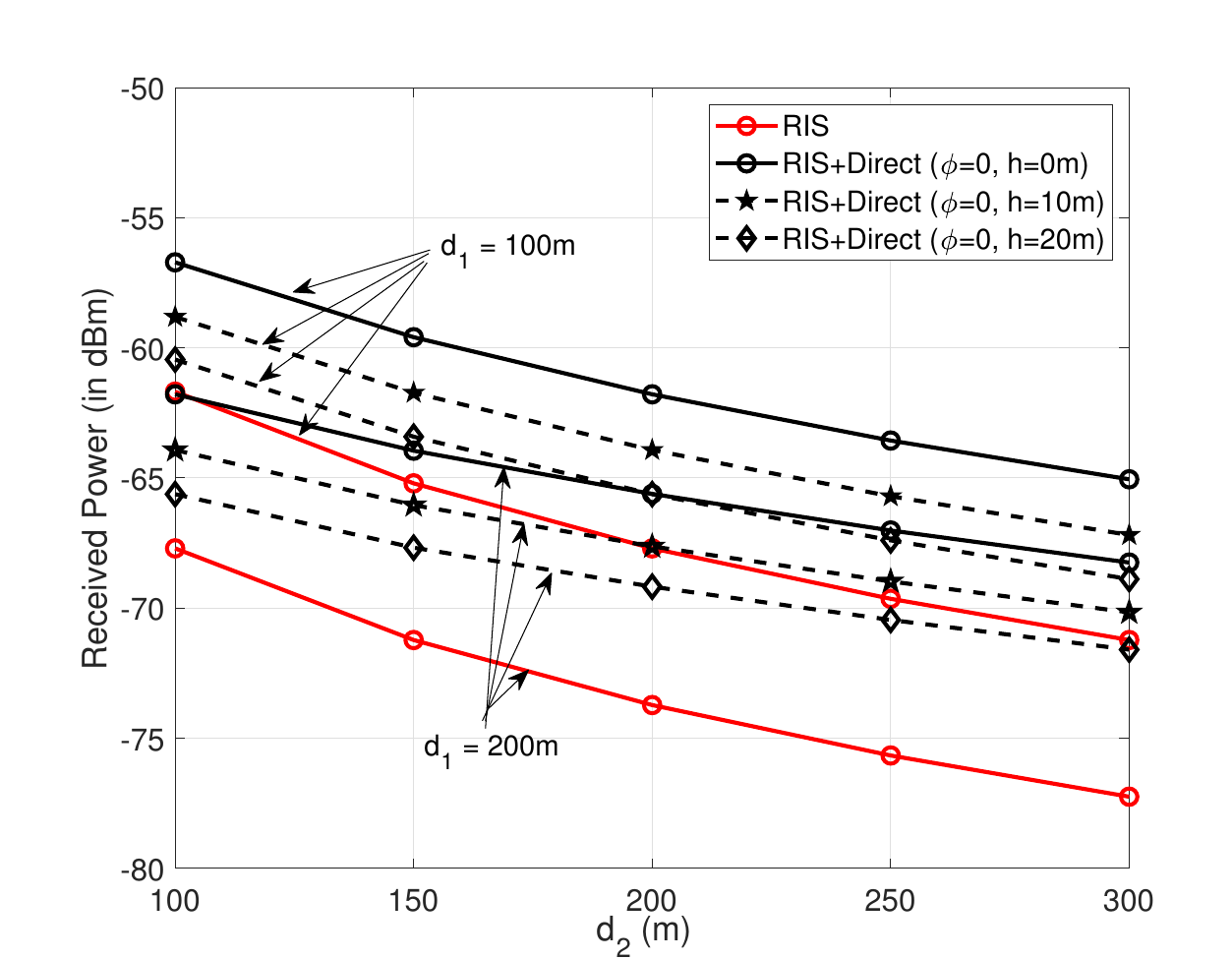}} \hspace{-2mm}
	\subfigure[Effect of RIS and transceiver elevation factors.]{\includegraphics[scale=0.35]{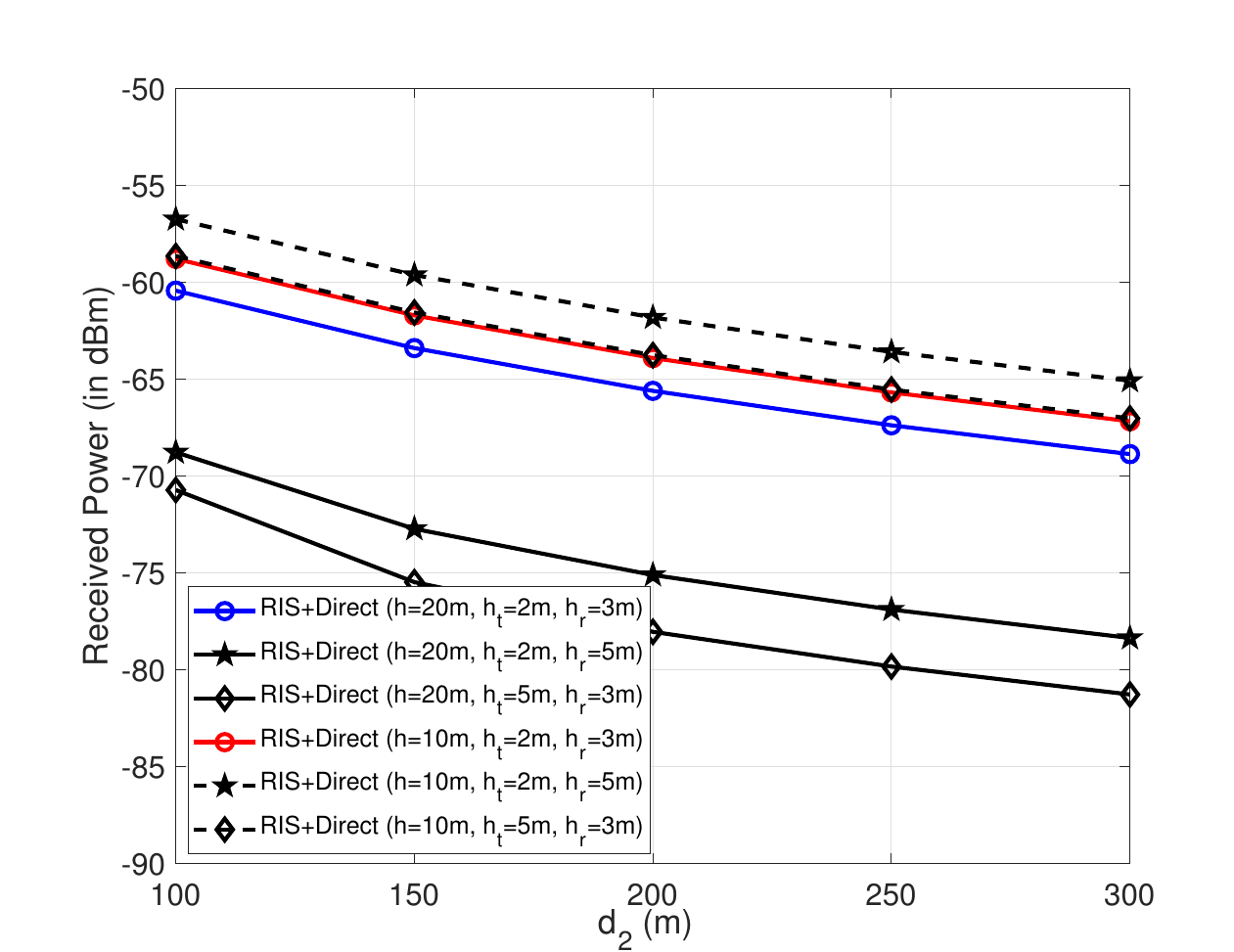}} \hspace{-2mm}
	\caption{Received power (dBm) for far-field broadcasting scenario with a transmit power of $10$\mbox{dBm}. }
	\label{fig:farf}
\end{figure*}

We plot the received power for the considered system for both near-field and far-field scenarios in Fig.~\ref{fig:NF_BF} and Fig.~\ref{fig:farf}, respectively. We take $P_t=10\mbox{dBm}$ such that path loss (PL) can be obtained from received power as $\text{PL (dB})=10-P_r(\text{dBm})$. We consider $G_t=G_r=21\mbox{dB}$, carrier frequency of  $10.5$\mbox{GHz}, and RIS configuration encompasses $M=102$, $N=100$ elements.  The transmitter and receiver angles are set to $\theta_t=\pi/4$ and $\theta_r=\pi/4$, respectively. Further, the transmitter and receiver phases are $\psi_t=\pi$ and $\psi_r=0$, respectively.

In  Fig.~\ref{fig:NF_BF}, we demonstrate the path loss for the near-field broadcasting considering different phase $\phi=0$ (perfect phase compensation) to $\phi=\frac {3\pi}{4}$ (minimum level of received power), varying height of RIS units from $h=0$ \mbox{m} to $h=15$ \mbox{m} at a fixed elevation of transceiver $h_t=2$ \mbox{m} and $h_r=3$ \mbox{m}. Note that   $h_t=h_r=h=0$ without direct link is the case considered in \cite{Tang_2021}. We have considered a shorter elevation of RIS  ($h\leq15$ \mbox{m}) due to the near-field scenario.   To demonstrate the near-field propagation, we fix the transmitter near the RIS ($d_1=2$ \mbox{m}) and increase the distance of the receiver from the RIS ($d_2=20$ \mbox{m} to $d_2=200$ \mbox{m}). The near-field criterion for the RIS is satisfied when either $d_1$ or $d_2$ is smaller than $71.4$\mbox{m}\cite{Tang_2021}. We maintain $d_1$ at $2$\mbox{m}, while allowing $d_2$ to vary from $20$\mbox{m} to $200$\mbox{m}.  The distance between transmitter and receiver is chosen to be $d=5(2h-h_{t}-h_{r})$ satisfying $d\gg(2h-h_{t}-h_{r})$.

Fig.~\ref{fig:NF_BF}(a) shows that the direct link impacts received signal power and the effective phase shift at each RIS element determines whether the signals are added constructively. For $\phi=0$, the combined received signal power is higher than both the direct signal and the reflected signal through RIS. However,  with the phase component $\phi=3\pi/4$, the received signal power is lower than individual components. The direct link improves $4$\mbox{dBm} in received power over extended distances. The phase $\phi=\pi/2$ reduces received power, consequently leading to an increased path loss. Fig.~\ref{fig:NF_BF}(b) demonstrates the impact of elevation of RIS $h$ on received signal power for fixed antenna heights. The figure shows that received power decreases with a change in height $h$. When $\phi=0$ and $h=0$, the received power is approximately $-22$ \mbox{dBm}. However, the received power reduces by $3$ \mbox{dBm} using the combined RIS and the direct link component with $h=10$ \mbox{m}. Further, when $h$ increases to  $15$ \mbox{m}, the received power decreases, resulting in a difference of approximately $4$ \mbox{dBm}. A similar trend can be observed in the received power for the phase $\phi=\pi/4$. It is important to note that the effect of RIS elevation on the received power is not significant when $\phi=0$.

Fig.~\ref{fig:farf} illustrates the relationship between received power and $d_2$ under far-field conditions, where both $d_1$ and $d_2$ exceed $71.4$\mbox{m}, satisfying the far-field condition. The RIS phase is fixed to $\phi=0$ with two different RIS elevations for $h$ are explored at $h=10$\mbox{m} and $h=20$\mbox{m}.   The transceiver elevation is  fixed at $h_t=2$\mbox{m} and $h_r=3$\mbox{m}. While $d_1$ is fixed at $100$\mbox{m} and $200$\mbox{m}, $d_2$ varies between $100$\mbox{m} and $300$\mbox{m}.

In Fig.~\ref{fig:farf}(a), it can be seen how the received power changes with elevation factors. There is a decrease in the received power of $6$\mbox{dBm} when  $d_1=100$\mbox{m} increases to $d_1=200$\mbox{m}. Further, the received power decreases when the RIS elevation factor  $h$ increases with a pronounced effect when  $\phi=0$. The decrease in received power is approximately $6$\mbox{dBm} when the elevation factor $h$ increases by $10$\mbox{m},  indicating a significant impact of RIS elevation on the received power in the far-field than the near-field scenario. In Fig.~\ref{fig:farf}(b), we show the impact of antenna heights on received signal power illustrating how the received power changes with $h_t$ and $h_r$ at  different $h$. The change in received signal power depends on the ratio of transmit and receive antenna heights to that of the RIS elevation $h$. As such, the received power decreases with antenna heights at the transmitter and receiver. 
When the elevations of both the transceiver and RIS are similar, the expression $h^{2}+h_{t}h_{r}-hh_{t}-hh_{r}\approx0$ suggests that elevation differences have little effect, and the effective signal power relies mainly on direct and RIS signals. This observation is evident in Fig.~\ref{fig:farf}(b), where changes in received signal power are more noticeable at a transceiver height of $h=20$\mbox{m} compared to $h=10$\mbox{m}.
\section{Conclusions}
We developed an analytical model for the path loss of a wireless system utilizing a multiple-element RIS module positioned above the ground surface with elevated transmitter and receiver configurations. Further, the direct link was integrated into the path loss model to enhance its applicability. Simplified analytical expressions were also presented for path loss under various configurations, including near-field and far-field scenarios. These expressions elucidated the impact of elevation factors on path loss, facilitating more accurate signal quality estimation at the receiver. Simulation results corroborated that accounting for elevated RIS modules and transceiver units could yield improved deployment strategies for RIS-based wireless systems. The path loss increases when the elevation factors accounting for the transceiver and RIS increase with almost perfect phase compensation at the RIS units.   The proposed analysis underscores the intricate interplay between spatial dynamics, phase configurations, and distance variations, offering valuable insights into optimizing the performance of RIS-enabled communication systems. Validation using experimentation is an excellent future scope of the proposed work. 
\section*{Appendix A: Far-Field}
Using \eqref{eq:main}, we can represent received power for RIS-assisted system as
\begin{flalign}\label{eq:two_ray_ris_5}
&P_r = P_t \big[\frac{\lambda}{4\pi}\big]^2\bigg|\frac{\sqrt{G_{t}G_{r}}}{d_{l}}   + \frac{\sqrt{G_r G_t G F_{}^{}(\theta_{t},\psi_{t})F_{}^{}(\theta_{r},\psi_{r}) d_x d_y}}{2 \sqrt{\pi}d_{1} d_{2}}  \nonumber\\&\sum_{m=1 - M/2}^{M/2} \sum_{n=1 - N/2}^{N/2}\Gamma_{n,m}  e^{\frac{j2\pi(d_{1}+d_{2}-\alpha)}{\lambda}}e^{\frac{-j2\pi (d_{1}+d_{2}-d_{l})}{\lambda}}\bigg| ^2 
\end{flalign}
where $\alpha=r_{n,m}^{t}+r_{n,m}^{r}$. The coordinates of transmit antenna can be expressed as $(d_{1}\sin{\theta_{t}}\cos{\psi_{t}},d_{1}\sin{\theta_{t}}\sin{\psi_{t}},d_{1}\cos{\theta_{t}}+h_{t})$. Thus, square of the distance between the transmitter to the $(m,n)$-th element $U_{n,m}$ is given as 
\begin{flalign}\label{eq:rt}
(r_{n,m}^{t})^{2} &= (d_{1}\sin{\theta_{t}}\cos{\psi_{t}}-(m-\frac{1}{2})d_{x})^{2}\nonumber \\&+(d_{1}\sin{\theta_{t}}\sin{\psi_{t}}-(n-\frac{1}{2})d_{y})^{2}\nonumber\\  &+(d_{1}\cos{\theta_{t}}+h_{t}-h)^{2}
\end{flalign}
We can approximate  $r_{n,m}^{t}$ in \eqref{eq:rt} to get 
\begin{flalign}
r_{n,m}^{t} & \approx d_{1}-\sin{\theta_{t}}\cos{\psi_{t}}(m-\frac{1}{2})d_{x}\nonumber \\&- \sin{\theta_{t}}\sin{\psi_{t}}(n-\frac{1}{2})d_{y}+(h-h_{t})\cos{\theta_{t}}
\end{flalign} 
Similarly, we can express  distance between the  $(m,n)$-th element $U_{n,m}$ to the receiver as
\begin{flalign}
r_{n,m}^{r}&\approx d_{2}-\sin{\theta_{r}}\cos{\psi_{r}}(m-\frac{1}{2})d_{x}\nonumber \\&-\sin{\theta_{r}}\sin{\psi_{r}}(n-\frac{1}{2})d_{y}+(h-h_{r})\cos{\theta_{r}}
\end{flalign}
Thus, the parameter $d_1+d_2-\alpha(=r_{n,m}^{t}+r_{n,m}^{r})$  becomes
\begin{flalign}\label{eq:dalpha}
d_{1}+d_{2}-\alpha&=(\sin{\theta_{t}}\cos{\psi_{t}}+\sin{\theta_{r}}\cos{\psi_{r}})(m-\frac{1}{2})d_{x}\nonumber \\&+(\sin{\theta_{t}}\sin{\psi_{t}}+\sin{\theta_{r}}\sin{\psi_{r}})(n-\frac{1}{2})d_{y}\nonumber \\&-(h-h_{t})\cos{\theta_{t}}-(h-h_{r})\cos{\theta_{r}}
\end{flalign}
Further, we can use the following identity:
\begin{flalign}\label{eq:identity}
&\sum_{m=1 - M/2}^{M/2} \sum_{n=1 - N/2}^{N/2} e^{\frac{j2\pi(d_{1}+d_{2}-\alpha)}{\lambda}} =\nonumber \\&M N e^{-\J\frac{2\pi}{\lambda}((h-h_{t})\cos{\theta_{t}}+(h-h_{r})\cos{\theta_{r}})} \nonumber\\&\frac{\sinc(\frac{M\pi}{\lambda}(\sin{\theta_{t}}\cos{\psi_{t}}+\sin{\theta_{r}}\cos{\psi_{r}})d_{x})}{\sinc(\frac{\pi}{\lambda}(\sin{\theta_{t}}\cos{\psi_{t}}+\sin{\theta_{r}}\cos{\psi_{r}})d_{x})}\nonumber \\ &\frac{\sinc(\frac{N\pi}{\lambda}(\sin{\theta_{t}}\sin{\psi_{t}}+\sin{\theta_{r}}\sin{\psi_{r}})d_{y})}{\sinc(\frac{\pi}{\lambda}(\sin{\theta_{t}}\sin{\psi_{t}}+\sin{\theta_{r}}\sin{\psi_{r}})d_{y})}
\end{flalign}
We can also express $e^{-j\Delta\phi}$ for small $\Delta\phi$:
\begin{flalign}\label{eq:two_ray_ris_9}
e^{-j\Delta\phi}\approx 1-j\Delta\phi=  1-j\frac{4\pi \big[h^{2}+h_{t}h_{r}-h h_{t}-h h_{r}\big]}{\lambda d}
\end{flalign}
Assuming all RIS units to share the same reflection coefficient, $\Gamma_{n,m}=Ae^{\J\phi}$, and using \eqref{eq:dalpha}, \eqref{eq:identity}, and \eqref{eq:two_ray_ris_9} in \eqref{eq:two_ray_ris_5}, we get
\begin{flalign}\label{eq:two_ray_ris_10}
&P_r = P_t \left[\frac{\lambda}{4\pi}\right]^2\Bigg|\frac{\sqrt{G_{t}G_{r}}}{d_{l}}  +\nonumber \\& M N Ae^{\J\phi}\frac{\sqrt{G_r G_t G F_{}^{}(\theta_{t},\psi_{t})F_{}^{}(\theta_{r},\psi_{r}) d_x d_y}}{2 \sqrt{\pi}d_{1} d_{2}}\nonumber\\&e^{-\J\frac{2\pi}{\lambda}((h-h_{t})\cos{\theta_{t}}+(h-h_{r})\cos{\theta_{r}})}  \nonumber\\&\frac{\sinc(\frac{M\pi}{\lambda}(\sin{\theta_{t}}\cos{\psi_{t}}+\sin{\theta_{r}}\cos{\psi_{r}})d_{x})}{\sinc(\frac{\pi}{\lambda}(\sin{\theta_{t}}\cos{\psi_{t}}+\sin{\theta_{r}}\cos{\psi_{r}})d_{x})}\nonumber \\&\frac{\sinc(\frac{N\pi}{\lambda}(\sin{\theta_{t}}\sin{\psi_{t}}+\sin{\theta_{r}}\sin{\psi_{r}})d_{y})}{\sinc(\frac{\pi}{\lambda}(\sin{\theta_{t}}\sin{\psi_{t}}+\sin{\theta_{r}}\sin{\psi_{r}})d_{y})} \nonumber \\&\bigg[1-j\frac{4\pi}{\lambda d} \big[h^{2}+h_{t}h_{r}-h h_{t}-h h_{r}\big]\bigg]\Bigg| ^2 
\end{flalign}
Simplifying \eqref{eq:two_ray_ris_10}, we get  \eqref{eq:far_field}, which completes the proof.

\section*{Appendix B: Near-Field Broadcasting}
Using the geometric optics and following analysis presented in \cite{Tang_2021} for the near-field broadcasting scenario without elevation factors, it can be seen that the signal received through RIS is equivalent to the signal received at the receiver after traveling a distance of $(d_{1}+d_{2})$ from the transmitter. Under this scenario,  we can use \eqref{eq:main} to express the received signal power as
\begin{flalign}\label{eq:two_ray_ris_14}
P_r = P_t \big[\frac{\lambda}{4\pi}\big]^2\bigg|\frac{\sqrt{G_{t}G_{r}}\sqrt{F_{direct}^{combine}}}{d_{l}} + \nonumber \\Ae^{\J\phi}\frac{\sqrt{G_r G_t}}{d_{1}+d_{2}} e^{\frac{-j2\pi(d_{1} + d_{2} - d_{l})}{\lambda}}\bigg| ^2 
\end{flalign}
Assuming $d\gg 2h-h_t-h_r$ (with $d_2$ or $d_1$ very small) and applying Taylor's series, we can approximate the path difference as $\Delta l= d_{1}+d_{2}-d_{l} \approx \frac{2}{d}\big[h^{2}+h_{t}h_{r}-h h_{t}-h h_{r}\big] $. Thus, \eqref{eq:two_ray_ris_14} becomes
\begin{flalign}\label{eq:two_ray_ris_15}
P_r = P_t \big[\frac{\lambda}{4\pi}\big]^2\bigg|\frac{\sqrt{G_{t}G_{r}}\sqrt{F_{direct}^{combine}}}{d_{l}} + Ae^{\J\phi}\frac{\sqrt{G_r G_t}}{d_{1}+d_{2}} \nonumber \\ \bigg[1-j\frac{4\pi}{\lambda d} \big[h^{2}+h_{t}h_{r}-h h_{t}-h h_{r}\big]\bigg]\bigg| ^2 
\end{flalign}
After straightforward simplification of \eqref{eq:two_ray_ris_15}, we can get \eqref{eq:two_ray_ris_16}, which completes the proof.
\bibliographystyle{IEEEtran}
\bibliography{Multi_RISE_full}

\end{document}